# Superconducting Superstructure for the TESLA Collider

J. Sekutowicz*, M. Ferrario#, C. Tang*

* DESY, MHF-SL, Hamburg, (Germany) - # INFN, Frascati, (Italy)

**Abstract**

We discuss the new layout of a cavity chain ( superstructure) allowing, we hope, significant cost reduction of the RF system of both linacs of the TESLA linear collider. The proposed scheme increases the fill factor and thus makes an effective gradient of an accelerator higher. We present mainly computations we have performed up to now and which encouraged us to order the copper model of the scheme, still keeping in mind that experiments with a beam will be necessary to prove if the proposed solution can be used for the acceleration.

# 1 INTRODUCTION

Recent results have shown that two main technical specifications: the accelerating gradient $E_{acc}$ and the quality factor $Q_o$, 25 MV/m and $5 \cdot 10^9$ respectively, are achievable for bulk niobium cavities [1, 2]. The R&D program at DESY, to establish superconducting technology for at least 25 MV/m, is continued in order to reach specifications more repetitively and to lower the cost of this technology. The essential part of the total investment is the cost of the RF-system, meant here as the sum of cost of accelerating structures with auxiliaries and cost of RF-power distribution system. To cut this cost more effort should be done to:

- decrease the number of RF components, like: fundamental mode (FM) couplers, HOM couplers, waveguides, circulators, waveguide transformers.., per unit length,
- increase the effective gradient $E_{eff}$ in the collider.

In the present TTF design there are: 1 FM coupler and 2 HOM couplers per 9-cell structure which is almost 1 m long. The consequence of such dense positioning of FM couplers is that the RF-power distribution system becomes complex and thus more expensive.

The effective accelerating gradient in both linacs will be low, only 17.8 MV/m, when cavities will be operated at 25 MV/m. There are two reasons for that: a too small fill factor and the unflatness of the accelerating field.

The fill factor, defined here as ratio:

$$\text{fill factor} \equiv \frac{\text{cavity active length}}{\text{cavity total length}}$$

has a low value of 0.75, resulting from the length of interconnections between cavities, which are at present $3\lambda/2$ long (see Fig. 1). This length has been chosen at the very beginning of the sc linear collider studies.

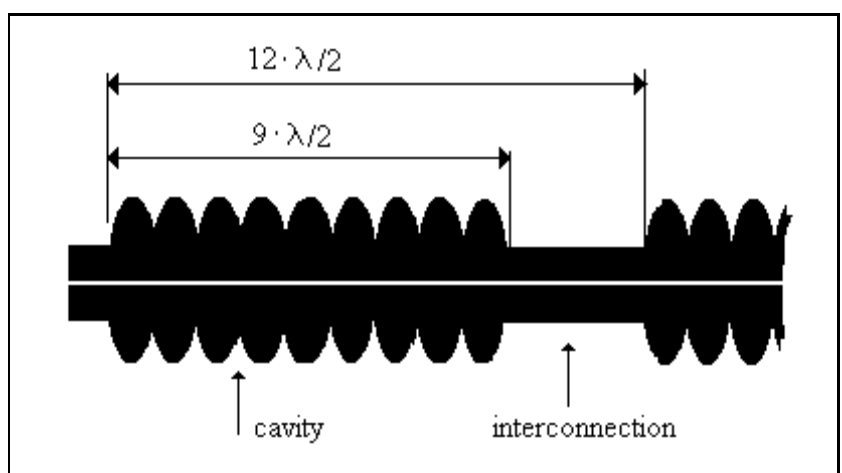

Fig. 1 The 9-cell cavity and the interconnection.

The arguments at that time were:
- good cavity separation for the accelerating mode and
- simplification in the phase adjustment.

The first argument will be discussed later but 7 km of passive length seems to be unjustified. The second argument is not valid any more since a 3-stub remote controlled waveguide transformer can be applied in the RF-input line of each cavity, to adjust both: the phase and the value of $Q_{ext}$, in order to get reflection-free operation.

The unflatness of the accelerating field within one structure is usually ~10 %. A typical field profile is shown in Fig. 2. For the accelerating π-mode, the sensitivity of the field amplitude $A_{cell}$ in an individual cell to the frequency error $\Delta f_{cell}$ of this cell, is given by the formula:

$$\Delta A_{cell} \sim \frac{(N)^2}{k_{cc}} \cdot \Delta f_{cell}$$

where N is the number of cells in the cavity and $k_{cc}$ is the cell-to-cell coupling.

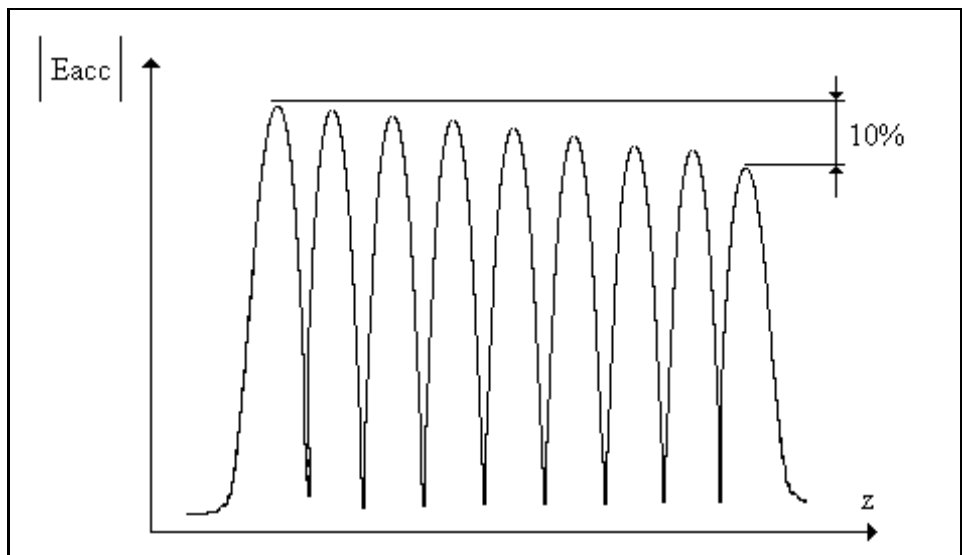

Fig. 2  An example of the field profile in a 9-cell TESLA cavity.

The experience with 20 TTF cavities showed that the current cavity design with N = 9 and $k_{cc}$ = 0.019 is almost at the limit. The design specification of the field unflatness below 5% is rather hard to obtain. Each chemical, thermal or mechanical cavity treatment tildes the field profile. This leads to a reduction of effective $E_{acc}$ since the achievable gradient is usually limited by the thermal break down in the cell with maximum amplitude.

A significant cost reduction can be done when the number of cells per structure increases. This is mainly due to lower number of RF components and less FM couplers per unit length. Unfortunately there are two fundamental limitations on N. First of all, the field profile, as it can be seen from the formula, becomes less stable, proportional to $N^2$. Secondly, the probability of trapping of parasitic resonances within the structure is higher. This is especially dangerous for sc cavities because even low (R/Q) parasitic modes can have finally big beam impedance due to high quality factor.

Since the length of interconnections seems to be oversized and simply increasing of N looks not very promising, we propose a different solution which is discussed in the next chapter.

## 2    SUPERSTRUCTURE

To overcome limitations on N and simultaneously to make interconnections shorter one may think to use the layout (superstructure) shown schematically in Fig. 3. The idea is to couple the cavities by short interconnections to enable an energy transfer from cavity to cavity instead of to separate them by a long interconnection. In this scheme, similar to the present design, HOM couplers are attached to interconnections and each cavity (sub-unit) is equipped with a tuner. This layout will allow to increase the number of cells fed by one FM coupler, avoiding the two limitations we discussed above. Both, the field flatness and the HOM damping, can be handled still at the sub-unit level.

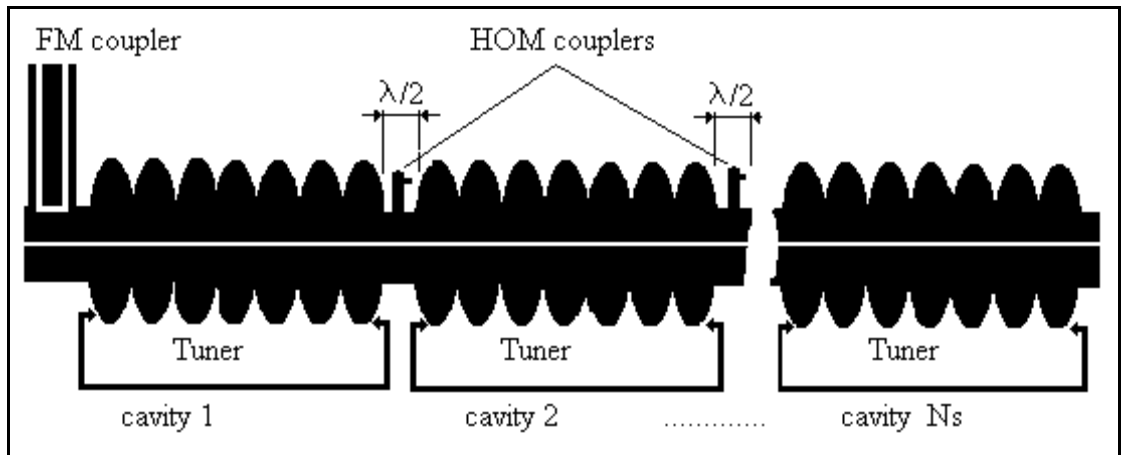

Fig. 3 Chain of $N_s$ cavities coupled by short interconnections (superstructure).

The length of the interconnection is chosen to be half of the wave length, $\lambda/2$. When N is an odd number, the $\pi$-0 mode ($\pi$ cell-to-cell phase advance and 0 structure-to-structure phase advance) can be used for the acceleration. As an example, the accelerating field profile of that mode in two neighboring cavities and in the interconnection is shown in Fig. 4. The expected coupling between sub-units depends now, since the length has been fixed, on the diameter of the interconnecting beam tube and on the field strength in the end cells. For the reasonable geometry of the interconnection this coupling is much smaller than the cell-to-cell coupling but using tuners for the frequency correction one can equalize the mean value of the field amplitude between sub-units ( not between cells within one sub-unit).

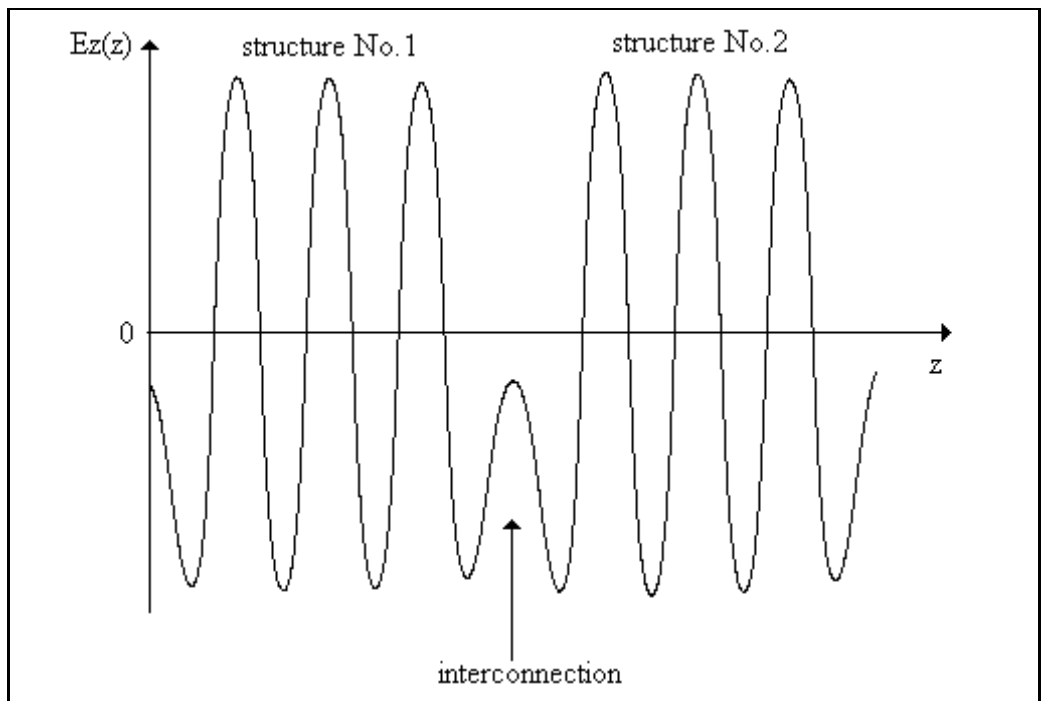

Fig. 4 An example of Eacc vs. z in 2 neighboring cavities excited in the $\pi$-0 mode.

## 3  COMPARISON OF TWO SUPERSTRUCTURES AND THE TTF CAVITY

The performance of the superstructure depends on the field profile stability within each sub-unit. There are two ways to make the accelerating field less sensitive to the cell frequency error (see previous formula):

- increasing coupling cell-to-cell, and /or
- reduction of number of cells per sub-unit.

The first proposed superstructure, made of four 9-cell cavities, had a more stable field in sub-units, as compared to the TTF cavity, due to the big mid iris diameter and almost 3 times bigger cell-to-cell coupling [3]. Unfortunately, the bigger aperture has some disadvantages: higher $E_{peak}/E_{acc}$ and $H_{peak}/E_{acc}$ and lower impedance (R/Q), than the present TTF design. These caused that the total improvement in effective $E_{acc}$ was rather small. Nevertheless, the proposed shape is better for alternative fabrication methods, like hydroforming or spinning, being still under development and which may in the future significantly reduce the investment cost [4, 5]. In addition, much lower transversal and longitudinal loss factors make this superstructure suitable for the acceleration of bunches with higher population of particles, like in the case of the muon collider.

It seems that the most probable future scenario for the energy upgrade of the TESLA collider, above 500 GeV, is the operation at higher accelerating gradient or/and making the collider longer [6]. This brought us to an alternative version of the superstructure [7], based on the TTF shape with modified end cells and reduced N from 9 to 7. As before, the superstructure is made of 4 sub-units. This version keeps $E_{peak}/E_{acc}$ and $H_{peak}/E_{acc}$ low as for the TTF cavity and makes operation above 25 MV/m more visible, since maximum electric and magnetic fields on the Nb wall are further from the theoretical limitations. Table 1 contains a list of parameters of both superstructures and the TTF cavity.

**Table 1  Comparison of two proposed superstructures and the TTF cavity**

| Parameter | unit | Big iris | Small iris | TTF cavity |
|---|---|---|---|---|
| mid / end iris radius | [mm] | 51/55 | 35/55 | 35/39 |
| N / $N_s$ | - | 9 / 4 | 7 / 4 | 9 / 1 |
| field instability factor, $N^2/k_{cc}$ | [ $10^3$ ] | 1.5 | 2.6 | 4.3 |
| sub-unit $(R/Q)_{cav}$ / m | [$\Omega$/m] | 668 | 911 | 995 |
| $E_{peak}/E_{acc}$ | - | 2.34 | 2.0 | 2.0 |
| $H_{peak}/E_{acc}$ | Oe/(MV/m) | 50.2 | 41.8 | 41.8 |
| $E1_{eff}$ (real flatn., $H_{peak}$ = 1065 Oe ) | [MV/m] | 18.4 | 21.2 | 17.8 |
| $E2_{eff}$ (real flatn., $E_{peak}$ = 50 MV/m ) | [MV/m] | 18.9 | 21.2 | 17.8 |

The two last rows of the table show $E_{eff}$ for the operation at 25 MV/m. In case of superstructures values: $E1_{eff}$ and $E2_{eff}$ are calculated with two limitations. The first one, for both values, results from the expected field unflatness, scaled from the value observed for the TTF cavities, proportional to the field instability factor, shown in the fourth row. The second limitation is $H_{peak}$ for $E1_{eff}$ and $E_{peak}$ for $E2_{eff}$. Here, the scaling is according to factors $H_{peak}/E_{acc}$ and $E_{peak}/E_{acc}$, respectively. Note, that the maximum improvement in the effective field is obtained for the superstructure based on a 7-cell sub-unit.

## 4  REFILLING OF CELLS AND THE BUNCH TO BUNCH ENERGY SPREAD

The most critical part of the numerical simulation is the calculation of the transient state and the bunch to bunch energy spread. Two codes: **HOMDYN** (beam dynamics and transients, see Appendix) and **LAPLACE** (transients only), showed that there is enough time to re-fill the cell's energy in the superstructure before the next bunch arrives [8, 9]. This result is rather not obvious since coupling between sub-units is very small.

As an example, the computed energy gain for the small iris superstructure, when it is operated at 25 MV/m, is shown in Fig. 5a, b.

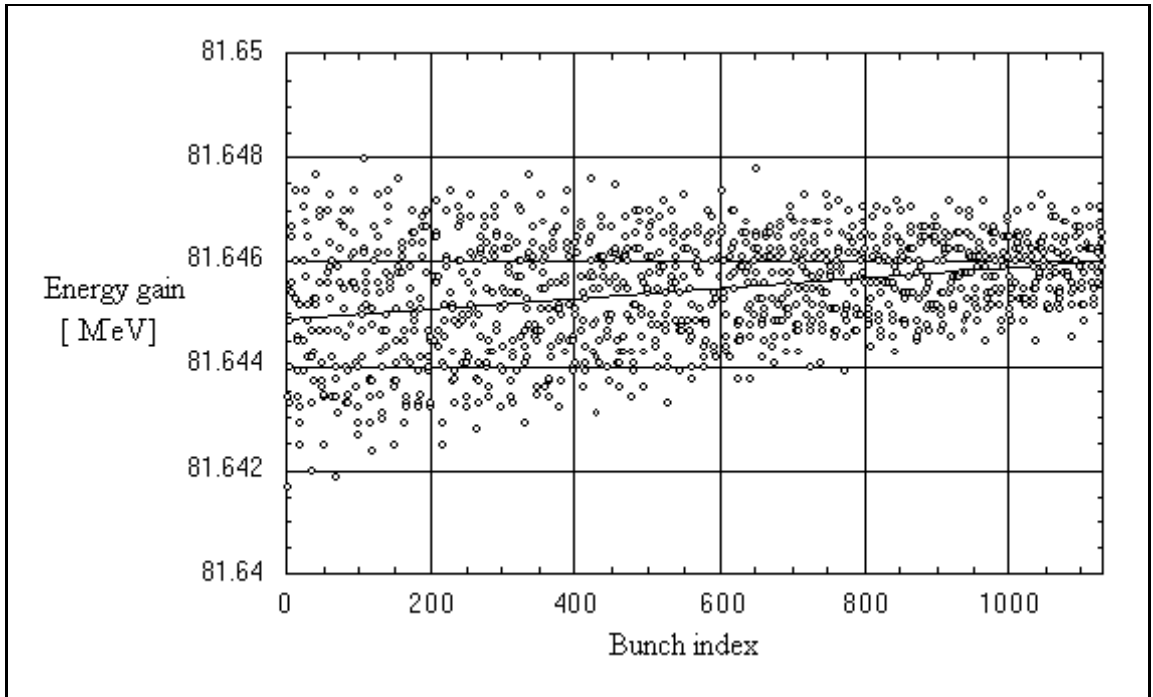

Fig. 5a. Energy gain for 1130 bunches accelerated with the small iris superstructure.

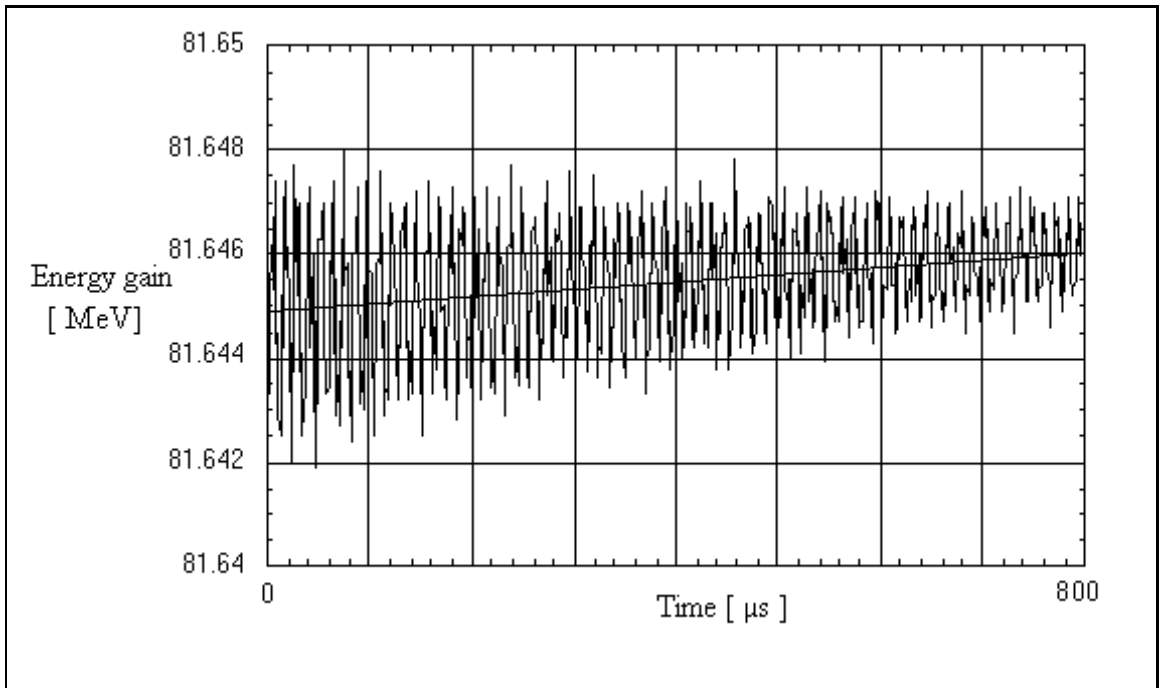

Fig. 5b. Energy gain vs. time.

The energy spread is mainly due to a small error in the injection time $t_o$ and to the interference of accelerating mode with mode $\pi-\pi/4$. The mean value of the energy (solid line in Fig. 5a) increases asymptotically. This indicates that the beam was injected few picoseconds too early and the accelerating voltage in the superstructure rises until the match condition is reached.
The oscillation of the bunch energy, better seen in Fig. 5b, has a small amplitude and the frequency $f = 80$ kHz, which equals to the difference between frequencies $f_{\pi-\pi/4}$ and $f_{\pi-0}$. The maximum energy spread for the whole train of 1130 is given for both superstructures in Table 2.

Table 2  Computed energy spread

|  | Big iris | Small iris |
|---|---|---|
| Energy spread | < 5 $10^{-4}$ | < 7 $10^{-5}$ |

## 5  DISCUSSION

In addition, to the improvement in the effective accelerating gradient, the number of FM and HOM couplers will be significantly reduced, if the proposed scheme can be used for the acceleration. Table 3 shows the total amount of couplers and tuners needed in the TESLA collider in two cases: when linac's layout is based on the current TTF cavity design and alternatively, when it is based on the small iris superstructure.

Table 3  Number of FM, HOM couplers and tuners

|  | TTF design | $N_s = 4$ |  |
|---|---|---|---|
| number of FM couplers | 19230 | 6181 | + |
| number of HOM couplers | 38460 | 24724 | + |
| power/FM coupler | 208 kW | 640 kW | - |
| number of tuners | 19230 | 24725 | - |

The needed number of FM couplers is reduced by a factor of 3. This has severe consequences for the cost of whole RF-system. When the diameter of interconnections is bigger than 114 mm all HOMs are above cutoff. Their field strength in the interconnections seems to be high enough for damping with HOM couplers attached at mid of interconnections. In that way each HOM coupler can be used to damp modes from two neighboring cavities. Such a damping scheme requires less HOM couplers than the TTF damping scheme.

The small iris superstructure based on a 7-cell sub-unit increases the total number of cavities by 22 %. This is an additional cost for 22 % more tuners and LHe vessels. Nevertheless the mentioned simplification in the RF system and the simplification in the cryostat construction will dominate and a total cost reduction can be expected.

The proposed layout is not yet proven experimentally. In the near future a copper model of the superstructure will be ordered. The RF-measurements on that model should help us to make a cross check with the computation we have done up to now for the superstructure in the superconducting and the normalconducting version.
We will be able to check on the copper model:

- tuning and field profile adjustment,
- transient state in individual cells,
- HOM damping scheme,
- coupling to FM coupler,
- influence of machining errors.

We won't be able to prove with this copper model the numerical simulation of the bunch-to-bunch energy spread. For that Nb prototype must be built and tested with the beam.

The power transfer by the FM coupler feeding superstructure is 640 kW. The new version of the FM coupler developed at DESY has been tested up to 1 MW for whole TESLA pulse length [10]. The limitation was due to the RF-power source. Since this version already overcame the power needed for operation of superstructure at 25 MV/m we do not expect here fundamental difficulties.

## ACKNOWLEDGMENTS

We would like to express our gratitude to R. Brinkman, D. Proch and the TESLA collaboration group for many helpful discussions.
## REFERENCES

[1] D. A. Edwards, "TESLA Test Facility Linac-Design Report", TESLA Rep. 95-01.

[2] B. Aune, D. Trines, "Results from the DESY TESLA Test Facility (TTF) Linac", Proc. of PAC'97, Vancouver, May 12-16,1997.

[3] J. Sekutowicz, "Superconducting Superstructure", TTF Meeting, Orsay, June 25-27,1997.

[4] H. Kaiser, private communication.

[5] V. Palmieri et al.., "Recent Experience with the Spinning of 1.5 GHz Seamless Copper Monocells", Proc. of $7^{th}$ Work. on SRF, Gif sur Yvette, 1995.

[6] R. Brinkmann, "Status of the Design for the TESLA Linear Collider", Proc.of PAC'95, Vol. 1 , Dallas, May 1995.

[7] J. Sekutowicz, M. Ferrario, C. Tang, "Superconducting Superstructure" , Proc.of LC'97, Zvenigorod, September 29- October 3, 1997.

[8] M. Ferrario et al. , "Multi-Bunch Energy Spread Induced by Beam Loading in a Standing Wave Structure" , Particle Accelerators, Vol. 52, 1996.

[9] J. Sekutowicz, "Transient State in Standing Wave Accelerating Structures", Particle Accelerators, Vol. 45, 1994.

[10] D. Proch, private communication.

[11] J. Sekutowicz, M. Ferrario, C. Tang, "Computations and Measurements of Transients in a Superstructure Pre-prototype", Proc. of the TTF Meeting, DESY, March 9-11, 1998, TESLA 98-06, p. 179.

[12] J. Sekutowicz, "2D FEM Code with Third Order Approximation for RF cavity computation", Proc. of the LINAC 94, Tsukuba, pp. 284.

# APPENDIX - BEAM LOADING COMPUTATIONS

We studied the superstructure-beam interaction by means of the code HOMDYN (see [8] and other references quoted there for a more detailed discussion). Originally the code was developed for single and multi-bunch dynamics computation in injectors devices, where transition from classical to relativistic dynamics takes place and space charge effects dominate the bunch transverse dynamics. Such a code relies on a simple self-consistent model that couples a current density description of beam evolution with the Maxwell equations in the normal modes expansion form. It takes into account single bunch space charge effects, beam loading of a long train of bunches, build-up effects of higher order modes and an on axis localized generator in order describe the cavity re-filling from bunch to bunch passage. The code is of course suitable for a fully relativistic beam dynamics computation, especially when transient fields excitation plays an important role. Several cross checks with other similar models, PIC's codes and recently with experimental measurement of transient fields excitation in a superstructure [11], allow us to conclude that our model is reliable.

We recall in this appendix the main equations of the model concerning the case under study, with some new features we added recently. We represent the electric field in the cavity as a sum of normal orthogonal modes:

$$\mathbf{E}(t,\mathbf{r}) = \sum_n \Re\mathrm{e}\left[A_n(t)\mathbf{e}_n(\mathbf{r})\right] \tag{1}$$

with complex amplitude

$$A_n(t) = \alpha_n(t)e^{i\omega_n t} = \frac{a_n(t)}{2}e^{i(\omega_n t + \varphi_n(t))} \tag{2}$$

where $a_n(t)$ is a real amplitude. The field form factors :

$$\mathbf{e}_n(\mathbf{r}) = \frac{e_n(\mathbf{r})}{i} \tag{3}$$

are any normalized solution of the Helmholtz equation, satisfying the boundary condition:

$$\hat{\mathbf{n}} \times e_n = 0 \tag{4}$$

on the cavity surface and the solenoidal condition:

$$\nabla \cdot e_n = 0 \tag{5}$$

within the cavity volume. They can be computed by standard finite differences codes (SUPERFISH, MAFIA, etc.), or, as in the present case, by a finite element code recently developed [12]. In the following we will restrict our attention to the on axis longitudinal electric field components of TM modes. The modes amplitude equations are:

$$\ddot{A}_n + \frac{\omega_n}{Q_n}\dot{A}_n + \omega_n^2 A_n = -\frac{1}{\varepsilon_o}\frac{d}{dt}\left[\int_V J(z,t) \cdot e_n^*(z)dv\right] \tag{6}$$

where as a driving current density we consider the superposition of two terms $J = J_g + J_b$. The term $J_g$ is a feeding sinusoidal current density, representing a point like power supply on the cavity axis located at $z_g$. The second term $J_b$ represents the beam current density. The loaded quality factor Q accounts for the cavity losses.

We have included the possibility to change the rf pulse rising time $\tau_g$, as discussed in [9], representing the power supply term as follows:

$$J_g(t, z_g) = \frac{J_g^o}{2i} \delta(z - z_g) \left(1 - e^{-\frac{t}{\tau_g}}\right) e^{i(\omega_1 t + \psi_1)} \quad (7)$$

where $J_g^o$ is the generator strength, $\omega_1$ and $\psi_1$ are the generator frequency and phase respectively.

The basic assumption in the description of the beam term consists in representing each bunch as a uniform charged cylinder, whose length L and radius R can vary under a self-similar evolution, i.e. keeping anyway uniform the charge distribution inside the bunch. Further details are reported in [8], we recall here that the beam current density term $J_b$ can be written for each bunch as follows:

$$J_b(t, z) = \frac{q \beta_{bar} c}{L} [\eta(z - z_t) - \eta(z - z_h)] \quad (8)$$

where q is the bunch charge, $\beta = v(t)/c$, $\eta$ is a step function and the indexes h, t refer to bunch head and tail positions respectively. The equations for the longitudinal motion of the bunch barycenter are simply:

$$\dot{\beta}_{bar} = \frac{e}{m_o c \gamma_{bar}^3} E_z(t, z_{bar}) \quad (9)$$

$$\dot{z}_{bar} = \beta_{bar} c \quad (10)$$

Substituting the definition (2) in the modes amplitude equations (6), under the slowly varying envelope (SVEA) approximation

$$\frac{d\alpha_n}{dt} \ll \omega_n \alpha_n \quad (11)$$

we can neglect the second order derivatives

$$\frac{d^2 \alpha_n}{dt^2} \ll \omega_n^2 \alpha_n \quad (12)$$

and we obtain a first order amplitude equation for each mode:

$$\dot{\alpha}_n + \frac{\omega_n}{2Q_n}\left(1 + \frac{i}{2Q_n}\right)\alpha_n = -\frac{1}{2\omega_n \varepsilon_o}\left(1 + \frac{i}{2Q_n}\right)\frac{d}{dt}\left[\int J(z,t) \cdot e_n^*(z) dz\right] e^{-i\omega_n t} \quad (13)$$

The SVEA approximation supposes small field perturbations produced by any single bunch, that add up to give an envelope of any field mode slowly varying on the time scale of its period T. Because the characteristic cavity reaction time is of the order of

$$\tau = \frac{2Q}{\omega} \gg T \quad (14)$$

we fulfill the SVEA hypothesis. This approximation allows to reduce the numerical and analytical computing time. The evolution of the field amplitude during the bunch to bunch interval is given by an analytical solution of equation (13) with $J_b=0$, which connects successive numerical integration applied during any bunch transit. Taking into account the generator feeding current (7), with a general initial condition $\alpha_n(t_o) = \alpha_n^o$, the analytical solution of (13) is:

$$\alpha_n(t) = K_n \left\{ \begin{array}{l} \dfrac{i\omega_1}{i\Omega_n + \dfrac{\omega_n}{2Q_n}\left(1 + \dfrac{i}{2Q_n}\right)} \left[ e^{-\left(i\Omega_n + \dfrac{\omega_n}{2Q_n}\left(1 + \dfrac{i}{2Q_n}\right)\right)(t-t_o)} - 1 \right] e^{i\Omega_n t} + \\ + \dfrac{g - i\omega_1}{i\Omega_n + \dfrac{\omega_n}{2Q_n}\left(1 + \dfrac{i}{2Q_n}\right) - g} \left[ e^{-\left(i\Omega_n + \dfrac{\omega_n}{2Q_n}\left(1 + \dfrac{i}{2Q_n}\right) - g\right)(t-t_o)} - 1 \right] e^{(i\Omega_n - g)t} \end{array} \right\} + \alpha_n^o e^{-\dfrac{\omega_n}{2Q_n}\left(1 + \dfrac{i}{2Q_n}\right)(t-t_o)}$$

(15)

where $\Omega_n = \omega_1 - \omega_n$, $g = \dfrac{1}{\tau_g}$, and

$$K_n = \dfrac{1}{4i\varepsilon_o \omega_n}\left(1 + \dfrac{i}{2Q_n}\right) J_g^o e^{i\psi_1} e_n(z_g) \tag{16}$$